\documentclass[11pt,a4paper]{article}
\pdfoutput=1
\usepackage{jheppub}
\usepackage{amsmath}
\usepackage{amssymb}
\usepackage{graphicx}
\usepackage{rotating}
\usepackage{color} 
\usepackage{multirow} 
\usepackage[T1]{fontenc} 
\usepackage{slashed}

\newcommand {\black} {\color{black}}

\newcommand {\red} {\color{red}}

\def\321times{SU(3) $\times$ SU(2) $\times$ U(1)}
\def\rpv{$\slashed{R}_p \:$}
\def\rpvm{\slashed{R}_p \:}
\def\vb#1{\vbox to #1 pt{}}
\newcommand{\AddrAHEP}{
  {\it AHEP Group, Instituto de F\'{\i}sica Corpuscular --
    C.S.I.C./Universitat de Val{\`e}ncia \\
    Edificio de Institutos de Paterna, Apartado 22085,
  E--46071 Val{\`e}ncia, Spain}}

\newcommand{\AddrOrsay}{%
Laboratoire de Physique Th\'eorique, CNRS -- UMR 8627, \\Universit\'e de Paris-Sud 11, F-91405 Orsay Cedex, France
}

\newcommand{\AddrTrieste}{%
International School for Advanced Studies (SISSA),\\
via Bonomea 265, Trieste, Italy
}

\author[a]{F. Bazzocchi}
\author[b]{S. Morisi}
\author[b]{E. Peinado}
\author[b]{J. W. F. Valle}
\author[c]{A. Vicente}

\affiliation[a]{\AddrTrieste}
\affiliation[b]{\AddrAHEP}
\affiliation[c]{\AddrOrsay}

\emailAdd{fbazzo@sissa.it}
\emailAdd{morisi@ific.uv.es}
\emailAdd{epeinado@ific.uv.es}
\emailAdd{valle@ific.uv.es}
\emailAdd{avelino.vicente@th.u-psud.fr}

\def\gsim{\raise0.3ex\hbox{$\;>$\kern-0.75em\raise-1.1ex\hbox{$\sim\;$}}}
\def\lsim{\raise0.3ex\hbox{$\;<$\kern-0.75em\raise-1.1ex\hbox{$\sim\;$}}}

\abstract{Bilinear R-parity violation (BRPV) provides the simplest
  intrinsically supersymmetric neutrino mass generation scheme. While
  neutrino mixing parameters can be probed in high energy
  accelerators, they are unfortunately not predicted by the
  theory. Here we propose a model based on the discrete flavor
  symmetry $A_4$ with a single R-parity violating parameter, leading
  to (i) correct Cabbibo mixing given by the Gatto-Sartori-Tonin
  formula, and a successful unification-like b-tau mass relation, and
  (ii) a correlation between the lepton mixing angles $\theta_{13}$
  and $\theta_{23}$ in agreement with recent neutrino oscillation
  data, as well as a (nearly) massless neutrino, leading to absence of
  neutrinoless double beta decay.}

\begin{document}


\title{\red Bilinear R-parity violation with flavor symmetry}

\keywords{supersymmetry; neutrino masses and mixing; R-parity violation}


\maketitle

\section{Introduction}
\label{intro}

Significant progress has recently been made at the Large Hadron
Collider (LHC), with the discovery of a new scalar state in the $125$
GeV mass region~\cite{CMStalk,ATLAStalk}. Although a conclusive
identification is still not possible, the properties of this new state
resemble very much those expected for the long-awaited Higgs boson.
This already constitutes one of the most important discoveries of
modern physics and represents an incredible success for a 50-year old
theory.

If this new scalar state were confirmed to be the Higgs
boson, we would know that the Standard Model (SM) is indeed the
correct effective description of elementary particles at least up to a
scale which we still ignore. Measuring the exact Higgs boson mass
would be crucial to know up to which scale the Higgs boson scalar
potential is stable, or in other words at which scale we should expect
new physics to emerge~\cite{EliasMiro:2011aa}.
The hierarchy problem associated to the Higgs boson mass has suggested
that new physics should appear around the TeV scale. Since the most
promising extension of the SM to address the hierarchy problem is
supersymmetry (SUSY), we expected that the stop or the gluino should
be around the corner. However the first searches up to $\sim 5
fb^{-1}$ at the LHC~\cite{cms-searches} have pushed the bounds on
squark and gluino masses beyond the TeV scale.  Even if the analysis
have been performed within specific frameworks, such as Constrained
Minimal Supersymmetric Standard Model (CMSSM) or minimal super-gravity
(MSUGRA), what the most recent results suggest is that if SUSY exists
one probably should be open minded as to how exactly it is
realized. Open issues in this regard are the precise mechanism of SUSY
breaking and whether R-parity is conserved. Indeed, supersymmetry may
well be broken by a non-gravitational messenger.  Similarly, one can
have supersymmetry without
R-parity~\cite{hall:1983id,ross:1984yg}. Hence the need to consider
alternative scenarios~\cite{Dreiner:2012wm} where, in addition, the
stringent bounds on the squark and gluino masses are
relaxed~\cite{Graham:2012th,Hanussek:2012eh}.

Apart from stabilizing the Higgs boson scalar potential, supersymmetry
could address other Standard Model puzzles for which new physics is
invoked. Among these we have that supersymmetry might explain the
origin of neutrino masses as well as cold dark matter.
Regarding the latter it has recently been shown that a relatively
light gravitino in the few GeV range can provide a perfectly valid and
interesting alternative in broken R-parity
models~\cite{Restrepo:2011rj}. Moreover, it provides a testable
minimal mechanism for the origin of neutrino
masses~\cite{hirsch:2004he}.
Regarding neutrinos it is well-known that bilinear R-parity violation
offers a simple way to generate neutrino masses in
supersymmetry~\cite{hirsch:2004he}. In its ``generic'' formulation the
model can not address issues associated to fermion mass hierarchies
and mixings, such as those of neutrinos.  
Both Abelian~\cite{BenHamo:1994bq} and
non-Abelian~\cite{Carone:1996nd} flavor symmetries have been used in
the literature to constrain the R-parity violating terms. In this
letter we propose a flavored version of bilinear R-parity violation.
The model has a single supersymmetric R-parity violating parameter
allowed by the flavor symmetry $A_4\times Z_2$, where $A_4$ is the
group of even permutations of four objects. This R-parity violating
term is used to generate neutrino masses as required by current
oscillation data, see \cite{Maltoni:2004ei}. We obtain predictions for
the charged fermion masses as well as for neutrinos.

The paper is organized as follows: in section \ref{SBRPV} we briefly
review neutrino mass generation through low-scale supersymmetry with
bilinear R-parity violation~\cite{diaz:1997xc} and in section
\ref{Tmodel} we extend it by implementing a discrete flavor
symmetry. In section \ref{nupheno} we present our results in the
neutrino sector, where we find a correlation between the lepton mixing
angles $\theta_{13}$ and $\theta_{23}$. In section \ref{scalpot} we
comment on the scalar potential and finally in section \ref{conclu} we
summarize the main predictions of the model.

\section{Bilinear R-parity violation}
\label{SBRPV}

Bilinear R-parity Violation \cite{hall:1983id,diaz:1997xc} is the
minimal extension of the Minimal Supersymmetric Standard Model (MSSM)
that incorporates lepton number violation, providing a simple way to
accommodate neutrino masses in supersymmetry. The superpotential is
\begin{equation}\label{brpv-superpotential}
W=W^{MSSM} + \epsilon_i \widehat{L}_i \widehat{H}_u .
\end{equation}
The three $\epsilon_i = (\epsilon_e,\epsilon_\mu,\epsilon_\tau)$ parameters
have dimensions of mass and explicitly break lepton number. Their
origin (and size) can be naturally explained in extended models where
the breaking of lepton number is spontaneous~\cite{Masiero:1990uj}. In
that sense, BRPV can be seen as an effective description of a more
general supersymmetric framework for lepton number violation. In any
case, the $\epsilon_i$ parameters are constrained to be small
($\epsilon_i \ll m_W$) in order to account for the small neutrino
masses. Furthermore, the presence of the new superpotential terms
implies new soft SUSY breaking terms as well
\begin{equation}\label{brpv-soft}
V_{soft}^{b\mbox{-}\rpvm} = B_{\epsilon_i} \tilde{L}_i H_u ,
\end{equation}
where the $B_{\epsilon_i}$ parameters have dimensions of mass
squared. The $\epsilon_i$ and $B_{\epsilon_i}$ couplings induce vacuum
expectation values (VEVs) for the sneutrinos, $\langle
\tilde{\nu}_L\rangle \equiv v_{L_i}$, proportional to the
$\epsilon_i$, hence small, as required (we assume $B_{\epsilon_i} = B
\, \epsilon_i$).

In the presence of BRPV couplings, neutrinos and neutralinos mix,
giving rise to neutrino masses
\cite{Hirsch:2000ef,Diaz:2003as,Chun:1999bq}. In the basis
$(\psi^0)^T=
(-i\tilde{B}^0,-i\tilde{W}_3^0,\widetilde{H}_d^0,\widetilde{H}_u^0,
\nu_{e}, \nu_{\mu}, \nu_{\tau} )$ the neutral fermion mass matrix
$M_N$ is given by
\begin{equation} \label{brpv-massF0}
M_N=\left(
\begin{array}{cc}
{\cal M}_{\chi^0}& m^T \cr \vb{20} m & 0 \cr
\end{array}
\right),
\end{equation}
where ${\cal M}_{\chi^0}$ is the usual neutralino mass matrix and
\begin{equation}
m=\left(
\begin{array}{cccc}
-\frac 12g^{\prime }v_{L_e} & \frac 12g v_{L_e} & 0 & \epsilon_e \cr \vb{18}
-\frac 12g^{\prime }v_{L_\mu} & \frac 12g v_{L_\mu} & 0 & \epsilon_\mu  \cr
\vb{18} -\frac 12g^{\prime }v_{L_\tau} & \frac 12g v_{L_\tau} & 0 & \epsilon_\tau
\cr
\end{array}
\right),
\end{equation}
is the matrix that characterizes the breaking of R-parity. Note that
its elements are suppressed with respect to those in ${\cal
  M}_{\chi^0}$ due to the smallness of the $\epsilon_i$
parameters. Therefore, the resulting $M_N$ matrix has a type-I seesaw
structure and the effective light neutrino mass matrix can be obtained
with the usual formula $m_\nu^{0} = - m \cdot {\cal M}_{\chi^0}^{-1} \cdot
m^T$, which can be expanded to give
\begin{equation} \label{meff-lambda-brpv}
\left( m_\nu^{0} \right)_{ij} = a^{(0)} \Lambda_i \Lambda_j ,
\end{equation}
where $a^{(0)}$ is a combination of SUSY parameters and 
\begin{equation}\label{Lambda}
\Lambda_i = \mu v_i + v_d \epsilon_i,
\end{equation}
are the so-called \emph{alignment parameters}. The projective form of
$m_\nu^{0}$ implies only one eigenvalue is non-zero. A natural choice
is to ascribe this eigenvalue to the atmospheric scale.
In this case the required solar mass scale, $\Delta m_{sol}^2 \ll
\Delta m_{atm}^2$, arises radiatively, at the 1-loop level, correcting
the tree-level neutrino mass matrix in Eq.~\eqref{meff-lambda-brpv}.
Detailed computations of the 1-loop contributions to the neutrino mass
matrix can be found in Refs.~\cite{Hirsch:2000ef,Diaz:2003as}. The
corrections are of the type
\begin{equation}\label{1loop}
\left( m_\nu^{1} \right)_{ij} \approx  a^{(1)} \Lambda_i \Lambda_j + b^{(1)} (\Lambda_i \epsilon_j + \Lambda_j \epsilon_i) + 
c^{(1)} \epsilon_i \epsilon_j ,
\end{equation}
where the coefficients $a^{(1)}, b^{(1)}, c^{(1)} $ are complicated
functions of the SUSY parameters. This generates a second non-zero
mass eigenstate associated with the solar scale, and the corresponding
mixing angle $\theta_{12}$.
Note that the neutrino mixing angles are determined as ratios of \rpv
parameters $\epsilon_i$ and $\Lambda_i$. 

Let us say a few words about the phenomenology of BRPV. The breaking
of R-parity has an immediate consequence at colliders: the LSP in no
longer stable and decays typically inside the detectors.
Since LSP decays and neutrino masses have a common origin, one can
show that ratios of LSP decay branching ratios correlate with the
neutrino mixing angles measured at low energies
\cite{Mukhopadhyaya:1998xj}.
This establishes a tight link which allows one to use neutrino
oscillation data to test the model at the LHC
see~e.~g.~\cite{DeCampos:2010yu}.

\section{The flavored BRpV model}
\label{Tmodel}

Let us consider the MSSM particle content extended with one extra
singlet superfield $\hat{S}$ and a $A_4 \times Z_2$ flavor symmetry
with the assignments given in table~\ref{model}.  The superfield
$\hat{S}$ is required in order to generate the $\mu$ term, and is the
only singlet under $A_4$.
On the other hand the $Z_2$ symmetry forbids all \rpv operators with
the only exception of the bilinear terms $LH_u$, while the quark and
charged lepton sectors are very similar to those
in~\cite{Morisi:2011pt}.
The assumption that all matter fields as well as the up and down Higgs
doublets are in triplet representations of $A_4$ reduces the different
BRPV parameters to only one.
\begin{table}[h!]
\begin{center}
\begin{tabular}{|c|c|c|c|c|c|c|c|c|}
\hline
 & $\,\hat{Q}\,$ &$\,\hat{u}^c\,$&$\,\hat{d}^c\,$&$\,\hat{L}\,$ & $\,\hat{e}^c\,$ & $\,\hat{H}_u\,$ & $\,\hat{H}_d\,$ & $\,\hat{S}\,$\\
\hline
$A_4$ & $3$& $3$&$3$& $3$& $3$& $3$& $3$& $1$\\
\hline
$Z_2$ &$+$ &$-$ & $+$& $-$& $-$& $-$& $+$& $-$\\
\hline
\end{tabular}\caption{The model assignments}\label{model}
\end{center}\end{table}

The superpotential of the model is
\begin{equation}\label{superpot}
\mathcal{W} = Y (\widehat{L}\,\widehat{e}^c )_{3}\,\widehat{H_d}\,+
               \,\epsilon\,\widehat{L}\,\widehat{H_u}\,
              +\lambda\,\widehat{H_u}\,\widehat{H_d}\,\hat{S}\,+m_S\,\hat{S}\,\hat{S}.
\end{equation}
Note that, due to the product rule $3\times 3 = 1+1'+1''+3_1+3_2$,
where $1,1',1''$ are different singlets of $A_4$ and $3_{1,2}$ are
different triplets, the assignment in table~\ref{model} allows for two
different contractions in the usual charged lepton Yukawa
interactions, compactly denoted by the first term in
eq.~(\ref{superpot}).
This leads to the couplings $Y_{\delta}|\epsilon_{ijk}^{\delta}|\,L_i
e^c_j H_{d_k}$ with $\delta=1,2$. For $\delta=1$ $(ijk)=
(123),(231),(312)$ and for $\delta=2$ $(ijk)= (213),(321),(132)$.
The resulting quarks and charged lepton mass matrices have the
form~\cite{Morisi:2011pt,Morisi:2009sc}
\begin{equation}\label{massf}
M_{f}=\left(
\begin{array}{ccc}
0 & y_1^f \langle H_3^f \rangle & y_2^f \langle H_2^f \rangle \\
y_2^f \langle H_3^f \rangle & 0 & y_1^f \langle H_1^f \rangle \\
y_1^f \langle H_2^f \rangle & y_2^f \langle H_1^f \rangle & 0
\end{array}
\right),~~~~~f=u,~d,~l,
\end{equation}
where $d$-type quarks and charged fermions $l$ couple to the same
Higgs.
Note that since all matter fields and Higgs scalars are in triplets of
$A_4$, the diagonal elements of the charged fermion mass matrices
vanish.
With the VEV alignment~\footnote{This VEV alignment will be justified
  in Sec. \ref{scalpot}, where it will be explicitly shown to be
  consistent with the minimization of the scalar potential of our
  model. However, note that the equality of the second and third
  entries in eqs. (\ref{alig-1})-(\ref{alig-3}) is an assumption
  required to obtain phenomenologically acceptable fermion mass
  matrices.}
\begin{eqnarray}
\langle H_u\rangle &=& \frac{1}{\sqrt{2}} (v_{u_1},v_{u_2},v_{u_3}) = v_{u_3} (r^u,-1,1) \label{alig-1} \\ 
\langle H_d\rangle &=& \frac{1}{\sqrt{2}} (v_{d_1},v_{d_2},v_{d_3}) =v_{d_3} (r^d,-1,1) \label{alig-2} \\ 
\langle \tilde{\nu}_L\rangle &=& \frac{1}{\sqrt{2}} (v_{L_e},v_{L_\mu},v_{L_\tau}) = v_{L_\tau}(a^\nu,-1,1), \label{alig-3}
\end{eqnarray}
we have in the charged fermion sector nine parameters, one of which
can be reabsorbed. These are used to fit nine masses and three mixing
angles, hence four predictions emerge~\cite{Morisi:2011pt}, given
below as eqs.~(\ref{eq:ql}) and (\ref{Uf}), in addition to
$V_{ub}=0=V_{cb}$. At this stage we have unmixed leptons and CP
conserved in the quark sector. Small nonzero $V_{ub}, V_{cb}$ can
arise by mixing with vector-like
quarks~\cite{prep:2012}\footnote{Vector-like quarks and their
  phenomenology have been widely studied in the literature. See for
  example the recent Refs.~\cite{Botella:2012ju,Okada:2012gy}.}.
To see this in more detail, let us rewrite the fermion mass matrix in eq. (\ref{massf}). With the VEV alignments in eqs. (\ref{alig-1})-(\ref{alig-2}), the mass matrix in eq. (\ref{massf}) can be rewriten as
\begin{equation}\label{massf2}
M_{f}=\left(
\begin{array}{ccc}
0 & a^f & -b^f \\
b^f & 0 & a^f r^f \\
-a^f &b^f r^f & 0
\end{array}
\right),
\end{equation}
where $a^f=y^f_1 v_3^f$, $b=y_2^f v_3^f$. From eq. (\ref{massf2}) we can see the mass matrix for the charged fermions has only three free parameters which can be written as functions of the charged fermion masses. Now we can consider the squared mass matrix for the charged fermions, $M_{f}M_{f}^T$,
\begin{equation}
M_{f}M_{f}^T\approx \left(
\begin{array}{ccc}
(b^f)^2 & -a^f b^fr^f & a^f b^fr^f  \\
-a^f b^fr^f & 
(a^f r^f)^2 & -a^f b^f \\
a^f b^fr^f & -a^f b^f & 
(b^f r_f)^2 
\end{array}
\right)\label{massf3}
\end{equation}
where we have assumed $a^f \ll b^f \ll r^f$ (see below).  The
invariants of this matrix give rise to three equations in terms of the
fermion masses. From these one can find the parameters, $a^f$, $b^f$
and $r^f$ as funtions of the charged fermion masses
as~\cite{Morisi:2009sc,Morisi:2011pt}
\begin{eqnarray}
a^f\approx \frac{m^f_2}{m^f_3}\sqrt{m^f_1 m^f_2}\label{sola}\\
b^f\approx\sqrt{m^f_1 m^f_2}\label{solb}\\
r^f\approx \frac{m^f_3}{\sqrt{m^f_1 m^f_2}}.
\label{solr}
\end{eqnarray}
From eqn. (\ref{solr}) we have the first prediction of the model, a quark-lepton mass relation:
\begin{equation}
\label{eq:ql}
\frac{m_\tau}{\sqrt{m_e m_\mu}}\approx\frac{m_b}{\sqrt{m_d m_s}},
\end{equation}
due to the equality $r^d=r^l$.  As discussed in
Ref.~\cite{Morisi:2011pt} such a formula works very well
experimentally and, in contrast to the well-known Georgi-Jarlskog
relation, does not arise from Clebsch Gordan coefficients, but follows
simply from the equality of the two functions $r^d(m_d,m_s,m_b)=
r^l(m_e,m_\mu,m_\tau)$. Moreover, it involves mass ratios, instead of
absolute masses, hence more stable from the renormalization viewpoint.

The second prediction is the Cabibbo angle, which follows from the
fact that the matrix in eq.~(\ref{massf3}) is diagonalized
by~\cite{Morisi:2009sc}
\begin{equation}\label{Uf}
U_f\approx\left(\begin{array}{ccc}1&\sqrt{\frac{m_1^f}{m_2^f}}&0\\-\sqrt{\frac{m_1^f}{m_2^f}}&1&0\\0&0&1\end{array}\right).
\end{equation}
This formula is simply the well-known Gatto-Sartori-Tonin relation
\cite{Gatto:1968ss}. Indeed from the matrix in eq.~(\ref{massf3}) one
obtains for instance the $V_{12}$ mixing as $V_{12}\sim a^f/(b^f r^f)
\sim \sqrt{{m_1^f}/{m_2^f}}$ which gives the famous Gatto-Sartori-Tonin
relation.
 
Let us now turn to the neutrino sector.  As already discussed in the
previous section, the tree-level neutrino mass matrix in
eq.~(\ref{meff-lambda-brpv}) has rank one. However it is
straightforward to show that with the VEV alignment in
eqs.~(\ref{alig-1}), (\ref{alig-2}) and (\ref{alig-3}) we have
\begin{equation}
\Lambda_\mu = -\Lambda_\tau\,,
\end{equation}
where the $\Lambda_i$ defined in (\ref{Lambda}) now take the form
\begin{equation}\label{Lambda2}
\Lambda_i = \mu v_{L_i} + v_{d_i} \epsilon,
\end{equation}
with $\mu = \lambda \langle S \rangle = \lambda \, v_s/\sqrt{2}$. Note
that the $\epsilon_i v_d$ contributions that characterize ``generic''
BRPV models described in the previous section, have now become
$\epsilon v_{d_i}$, where $i = (1,2,3)$, with a single bilinear
$\epsilon$ parameter, due to the flavor symmetry.

Once the 1-loop corrections are included we have, $$m_\nu=
m_\nu^0+m_\nu^1.$$ Using eqs. \eqref{meff-lambda-brpv} and
\eqref{1loop} and imposing the VEV alignment in eqs. \eqref{alig-1},
\eqref{alig-2} and \eqref{alig-3}, we find the resulting neutrino mass
matrix
\begin{equation}\label{mnu-param}
m_\nu=\left(
\begin{array}{ccc}
 c+\alpha  (2b+\alpha a) & c+b (\alpha -1)-\alpha a  & b+c+\alpha(a+b) \\
 c+b (\alpha -1)-\alpha a  & a-2b+c & c-a \\
 b+c+\alpha(a+b)  & c-a & a+2b+c
\end{array}
\right) \, ,
\end{equation}
where the following definitions have been made
\begin{eqnarray}
\Lambda_e &=& \alpha \Lambda \\
\Lambda_\tau &=& - \Lambda_\mu = \Lambda \\
a &=& \left( a^{(0)}+a^{(1)} \right) \Lambda^2 \\
b &=& b^{(1)} \Lambda \epsilon \\
c &=& c^{(1)} \epsilon^2
\end{eqnarray}
Note that the 1-loop contributions are dis-aligned with respect to the
tree-level one, with the tree-level degeneracy lifted by radiative
corrections. In the limit $\alpha = 0$ the neutrino mass matrix has
the massless eigenvector which corresponds to $(2,-1,-1)^T$, called
tri-bimaximal-$1$ for instance in Ref.~\cite{Antusch:2011ic}.
In the limit of $\alpha=b=0$ the spectrum is tri-bimaximal.

Finally, in the limit $b = c = 0$ one recovers the tree-level mass
matrix $m_\nu^0$. This matrix has rank one, and thus only one
eigenvalue is non-zero, $m_{\nu_3} = a |\vec \Lambda|^2 = a (2 +
\alpha^2)$. The associated eigenvector lies along the direction
$(\alpha,1,-1)$. Although there are corrections from the charged
lepton sector and from the 1-loop contributions, the condition
$|\alpha| \ll 1$ ensures a small $\theta_{13}$ value~\footnote{In our
  numerical analysis we have found that $|\alpha| \sim 0.2$ leads to
  $\theta_{13}$ in the observed range.}. Similarly, one expects the
hierarchy $b,c \ll a$, since $b$ and $c$ are generated at the 1-loop
level, whereas $a$ is a tree-level parameter. This naturally implies
$m_{\nu_2} \ll m_{\nu_3}$.

In conclusion, the neutrino mass spectrum is compatible with normal
hierarchy, with a radiatively induced solar scale. The solar and
atmospheric mass square differences as well as the solar mixing angle
can be fitted as shown explicitly in Ref.\,\cite{Hirsch:2000ef}.

\section{Large $\theta_{13}$ and deviations from maximal atmospheric mixing}
\label{nupheno}

\begin{figure}[!h]
\centering
\includegraphics[width=0.49\linewidth]{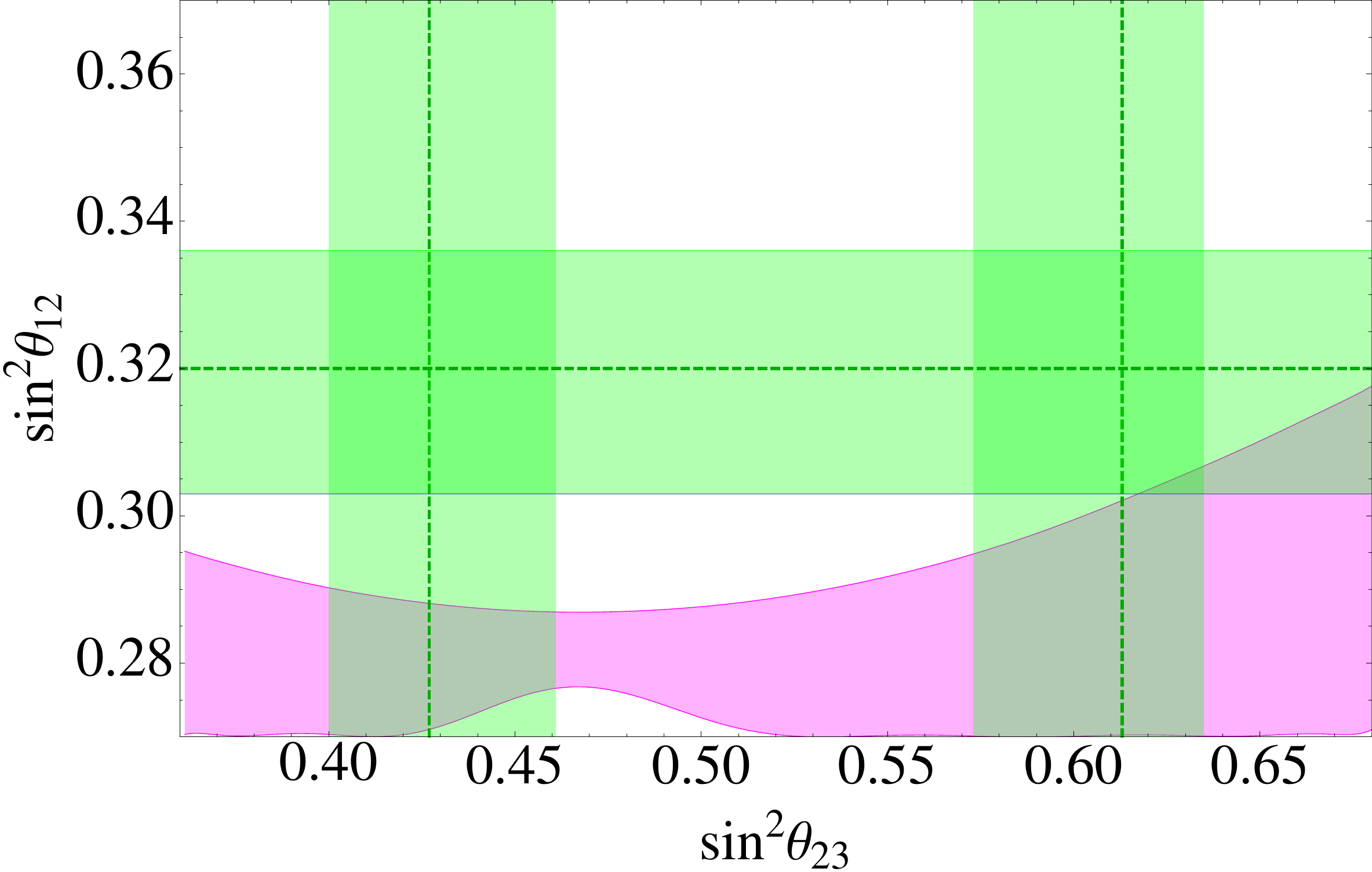}
\includegraphics[width=0.49\linewidth]{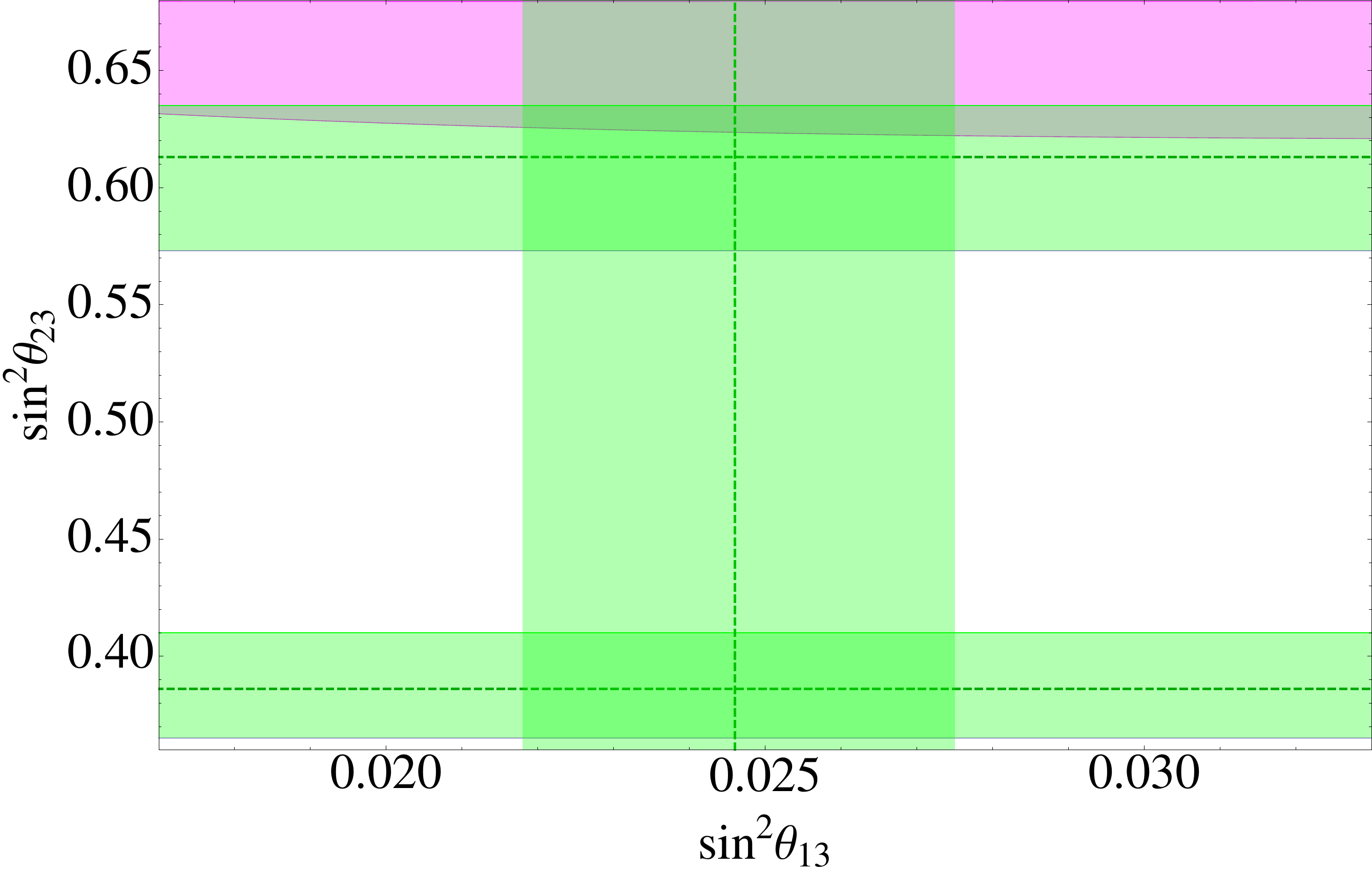}
\caption{ Left: Predicted correlation between the atmospheric and
  solar mixing angle taking the $3\sigma$ oscillation parameter ranges
  from the global fit of Ref.~\cite{Tortola:2012te}. The dashed lines
  represent the best-fit while the green and blue bands correspond to
  the $1\sigma$ range. As can be seen, there is a region consistent 
  at $1\sigma$ with the global fit of oscillation parameters.
  Right: The allowed range of reactor and atmospheric mixing angles,
  taking the $1\sigma$ range for the solar mixing angle. The dashed
  lines correspond to the best fit values, while the straight bands
  correspond to the $1\sigma$ ranges. As can be seen, maximal
  atmospheric mixing is excluded at $1\sigma$.}
\label{correlation1}
\end{figure}

\begin{figure}[!h]
\centering
 \includegraphics[width=0.49\linewidth]{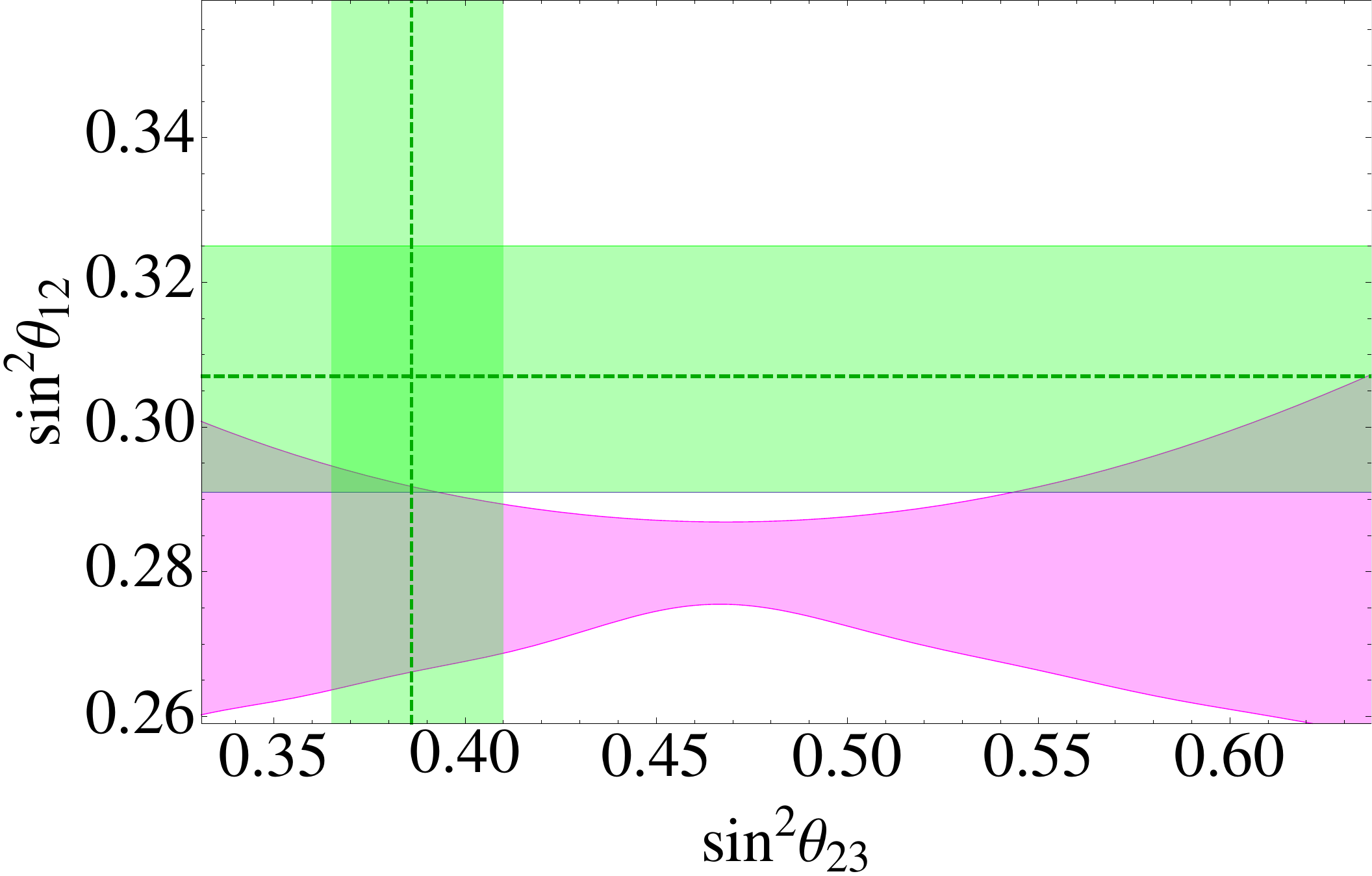}
 \includegraphics[width=0.49\linewidth]{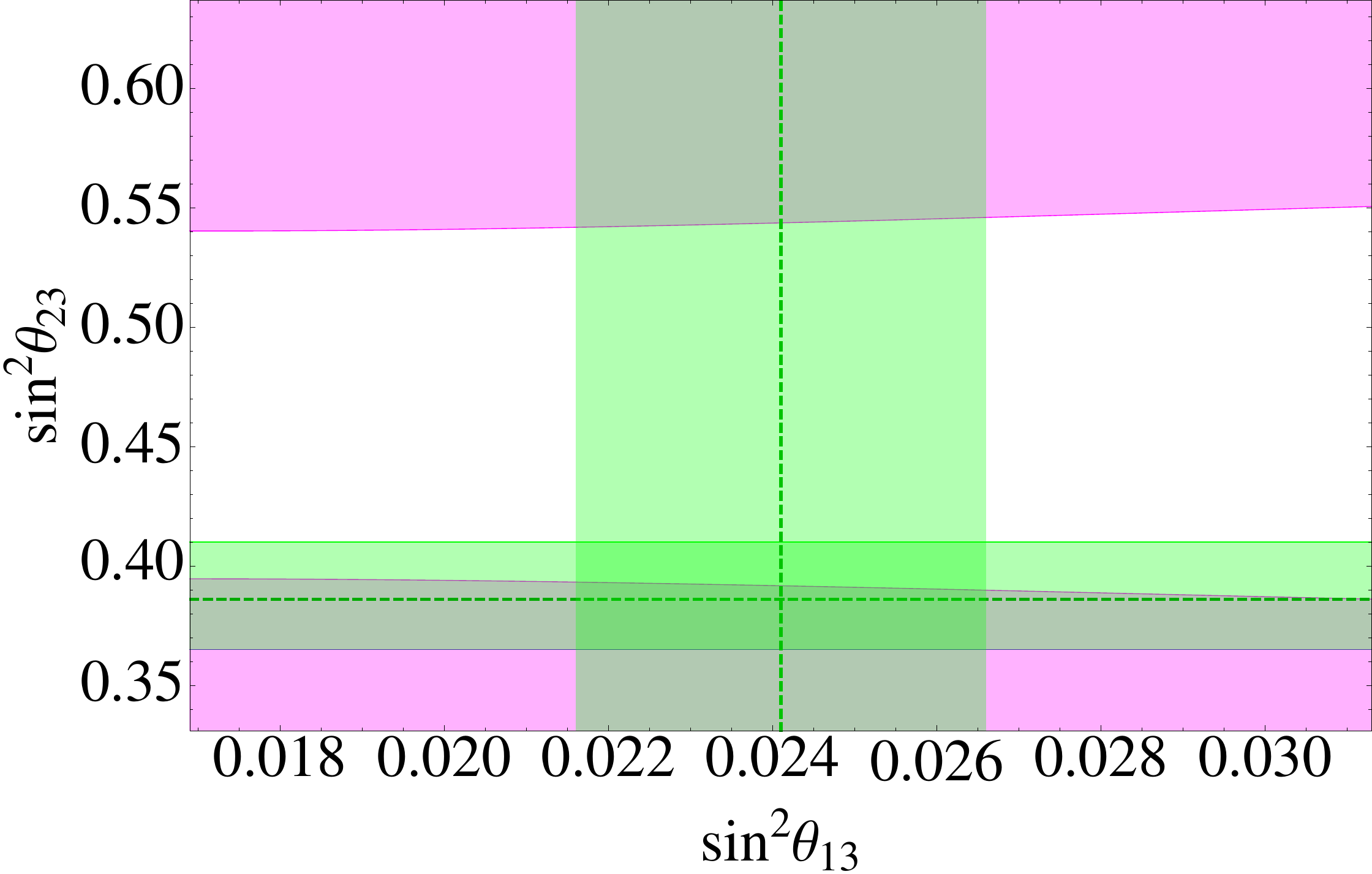}
\caption{ Same as Fig.~\ref{correlation1} for the analysis in
  Ref.~\cite{Fogli:2012ua}.}
\label{correlation2}
\end{figure}

\begin{figure}[!h]
\centering
 \includegraphics[width=0.49\linewidth]{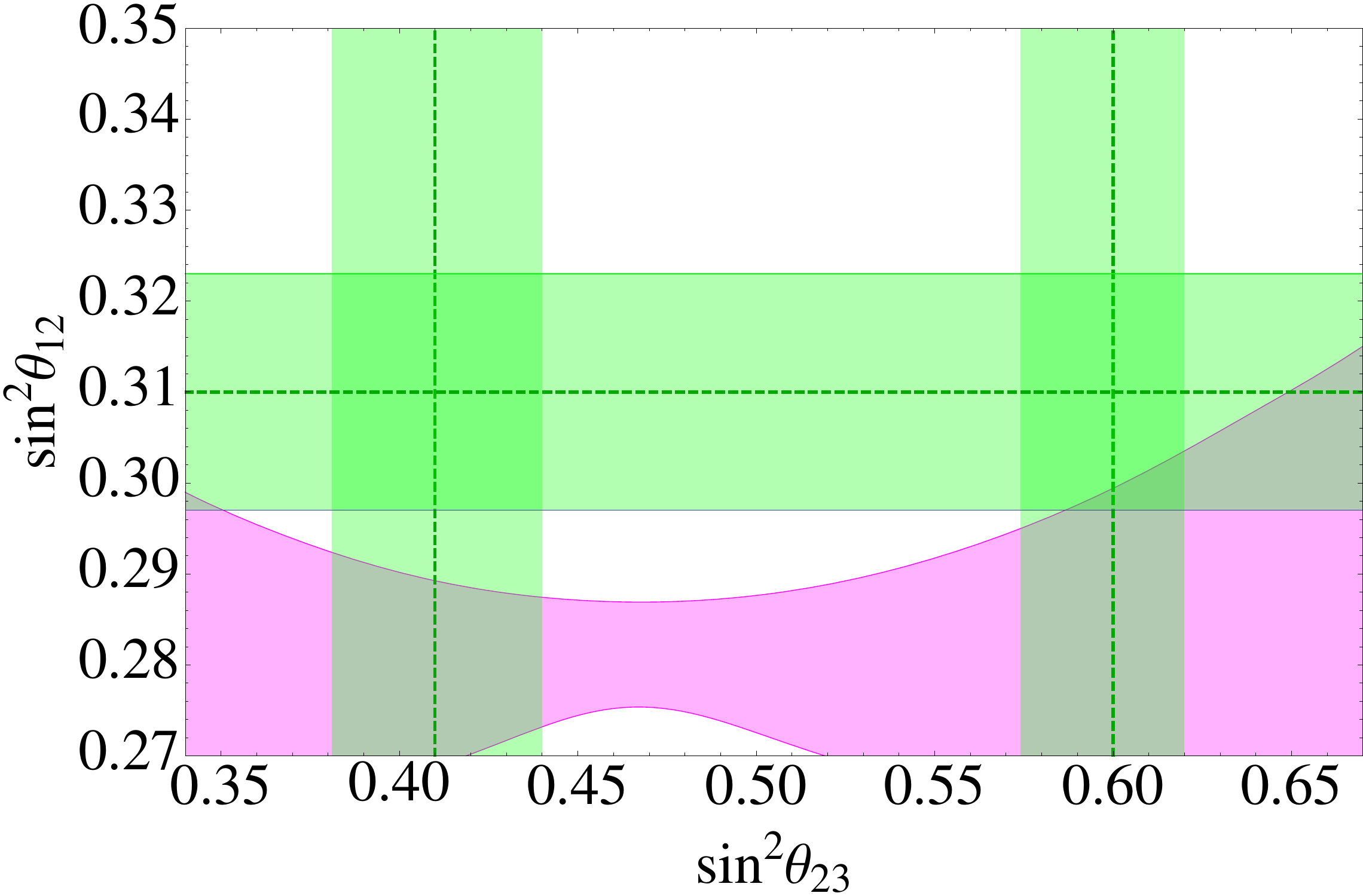}
 \includegraphics[width=0.49\linewidth]{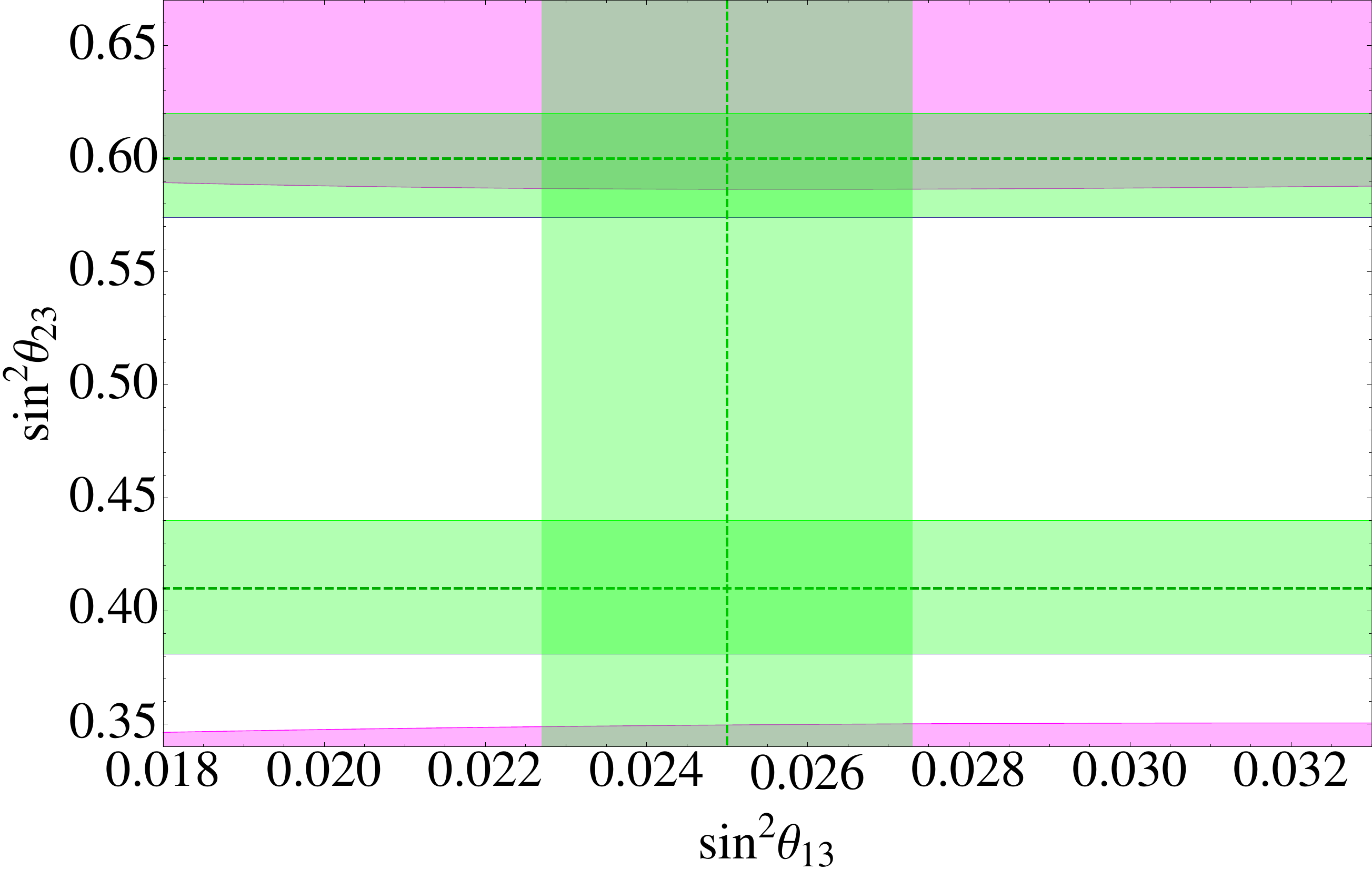}
\caption{ Same as Fig.~\ref{correlation1} for the analysis in
  Ref.~\cite{Gonzalez}.}
\label{correlation3}
\end{figure}

In the CP-conserving case, the neutrino mass matrix in
Eq. (\ref{mnu-param}) is characterized by $4$ free parameters, for $6$
observables in total, three masses and three mixing angles, therefore
two predictions can be obtained~\footnote{Note that neutrino mixing
  receives additional corrections from the charged lepton
  sector. However, these do not involve any additional
  parameter. Although relatively small, they have been taken into
  account in our numerical analysis.}. The first one is the mass of
the lightest neutrino, $m_{\nu_1}=0$, since the matrix in
Eq.~(\ref{mnu-param}) has a null eigenvalue (this state gets a
negligibly tiny mass once 2-loop contributions are included).
The second prediction is a correlation among the neutrino mixing
angles, which we determine for the recent global fits taking into
account the latest experimental data presented at the recent Neutrino
2012 conference.
The results of our analysis of the parameter space corresponding to
the global oscillation fits in Forero et al ~\cite{Tortola:2012te},
Fogli et al ~\cite{Fogli:2012ua} and Gonzalez-Garcia et
al~\cite{Gonzalez} are given in figures \ref{correlation1},
\ref{correlation2}, and \ref{correlation3}, respectively.

From the presented correlations between the atmospheric and solar
mixing angle, as well as the resulting allowed ranges of the reactor
and atmospheric angle, it is clear that our model restricts the
oscillation parameters in a non-trivial way, however consistent with
the $1\sigma$ ranges for the neutrino oscillation parameters given by
all global fits, in particular with the ``large'' reactor mixing angle
and non-maximal atmospheric mixing hinted by the most recent
oscillation data~\footnote{The main difference is the presence of two
  allowed octants for the atmospheric mixing in the analyses of
  \cite{Tortola:2012te,Gonzalez} while only one octant is present in
  the Bari group analysis, since it is more strongly preferred in that
  case~\cite{Fogli:2012ua}.}. At $2~\sigma$, predictions in our BRPV
model become very weak.

\section{Scalar potential and spectrum}
\label{scalpot}

The presence of additional Higgs doublets is a common feature to many
flavor models and leads to a complicated structure of the scalar
potential. Therefore, though non-trivial, the determination of a
phenomenologically viable minimum of the potential is of fundamental
importance. The scalar potential can be split as
\begin{equation}
V = V_F + V_D + V_{\text{soft}}^{A_4} + V_{\text{soft}}^{\slashed{A_4}}
\end{equation}
where $V_F$ and $V_D$ are the usual F- and D-terms and $V_{\text{soft}}^{A_4}$, defined as
\begin{eqnarray}
V_{\text{soft}}^{A_4} = \, && T_{Y_1} ( \tilde{L} \tilde{e}^*)_{3_1} H_d  + T_{Y_2} ( \tilde{L} \tilde{e}^*)_{3_2} H_d
 + T_\lambda H_{d} H_{u} S \nonumber \\ 
&& + B_\epsilon \tilde{L} H_{u} - B_{m_S} S^{2} + m_s^2 |S|^2+\mbox{h.c.}
\end{eqnarray} 
contains soft SUSY breaking terms that preserve the flavor
symmetry\footnote{For the sake of clarity we do not specify here the
  $A_4$ contractions. We also omit the soft gaugino masses, which of
  course preserve $A_4 \times Z_2$.}. In addition, we allow for the
existence of an additional piece, $V_{\text{soft}}^{\slashed{A_4}}$,
with terms that break softly both SUSY and $A_4$ but preserve
$Z_2$. The presence of this piece is necessary in order to obtain a
realistic spectrum. The $A_4 \times Z_2$ flavor symmetry leads to
accidental continuous symmetries in the scalar potential which, after
they get spontaneously broken by the corresponding VEVs, imply the
existence of additional Goldstone bosons. These massless states couple
to the gauge bosons thus the explicit breaking of those continuous
symmetries is required from a phenomenological point of
view\footnote{We do not attempt to provide a complete explanation about the origin of the terms in the scalar potential, nor about the hierarchy among them. On general grounds one expects that, for generic dynamics in the hidden sector responsible for supersymmetry breaking, the flavor symmetry is not respected. That is the reason why we allow for $A_4$ breaking terms in the soft SUSY breaking scalar potential.}.

We have considered the following soft breaking of $A_4$
\begin{equation}
V_{\text{soft}}^{\slashed{A_4}} = \sum_{ij} m_{H_{d_i} H_{d_j}}^2 H_{d_i} H_{d_j}^* + \sum_{ij} m_{H_{u_i} H_{u_j}}^2 H_{u_i} H_{u_j}^* + \sum_{ij} m_{l_i l_j}^2 \tilde{L}_{i} \tilde{L}_{j}^* + \sum_{ij} m_{e_i e_j}^2 \tilde{e}_{i} \tilde{e}_{j}^* 
\end{equation} 
Note that $A_4$ would enforce $m_{H_{d_i} H_{d_j}}^2 = m_{H_{u_i}
  H_{u_j}}^2 = m_{l_i l_j}^2 = m_{e_i e_j}^2 = 0$ for $i \ne j$ and
the equality of the soft masses for all elements in the same $A_4$
triplet. In our analysis we have broken those relations explicitly and
found minima of the scalar potential with realistic spectra in the
scalar, pseudoscalar and charged scalar sectors\footnote{In principle,
  additional terms of the type $m_{l_i H_{u_j}}^2$ are allowed in
  $V_{\text{soft}}^{\slashed{A_4}}$. However, the presence of such
  terms would introduce additional sources of lepton number violation
  and destabilize the required vacuum alignment.}.

The minimization of the scalar potential must also lead to the
required vacuum alignment, see equations \eqref{alig-1},
\eqref{alig-2} and \eqref{alig-3}. That restricts the allowed
parameter space. In particular, the conditions $|\alpha^u| \ll |r^u|$
and $|\alpha^d| \ll |r^d|$ can be naturally fulfilled with large soft
masses for the second and third generations of Higgs doublets in the
$(1-1000 \, \text{TeV})^2$ range (see below). That can be easily seen
from the tadpole equations. For example, for $v_{d_2}$ one finds
\begin{eqnarray}
\frac{\partial V}{\partial v_{d_2}} &=& \frac{1}{8} \Big(4 m_{H_{d_1} H_{d_2}}^2 v_{d_1} +8 m_{H_{d_2}}^2 v_{d_2} +4 m_{H_{d_2} H_{d_3}}^2 v_{d_3}+(g_{1}^{2} +g_2^2) (|\vec v_d|^2 - |\vec v_u|^2 + |\vec v_L|^2)v_{d_2} \nonumber \\  
 &&+4 |\lambda|^2 (v_{d_2} v_s^2 + v_{d_2} v_{u_2}^2 + v_{d_1} v_{u_1} v_{u_2} + v_{d_3} v_{u_2} v_{u_3}) \nonumber \\
 &&-4 \sqrt{2} v_s (2 v_{u_2} \: \text{Re} \left(\lambda m_S^*\right)  + v_{L_2} \: \text{Re} \left(\lambda^* \epsilon\right) + v_{u_2} {\: \text{Re}\left(T_{\lambda}\right)} \Big) = 0 \label{tadvd2}
\end{eqnarray}
Here we have defined $m_{H_{d_2}}^2 \equiv m_{H_{d_2} H_{d_2}}^2$,
$|\vec v_d|^2 \equiv \sum_i v_{d_i}^2$, $|\vec v_u|^2 \equiv \sum_i
v_{u_i}^2$ and $|\vec v_L|^2 \equiv \sum_i v_{L_i}^2$. Neglecting
small \rpv contributions and assuming the aforementioned VEV hierarchy
and CP conservation, equation \eqref{tadvd2} can be solved to give the
simple estimate
\begin{equation} \label{tadvd2-2}
v_{d_2} \simeq - \frac{4 v_{d_1} m_{Hd_1 Hd_2}^2}{(g_1^2+g_2^2)(v_{d_1}^2-v_{u_1}^2) + 8 m_{H_{d_2}}^2 - 4 m_{Hd_2 Hd_3}^2 + 4 \lambda^2 v_s^2}\approx\frac{\sqrt{m_d m_s} }{m_b}v_{d_1},
\end{equation}
where the last equality is obtained from eqns. (\ref{alig-2}) and
(\ref{solr}). Thus $|v_{d_{2}}| \ll |v_{d_1}|$ is obtained if
$m_{H_{d_2}}^2 \gg m_{H_{d_1} H_{d_2}}^2 , m_{H_{d_2} H_{d_3}}^2 \sim
m_{\text{SUSY}}^2$. Similar tadpole equations can be found for
$v_{d_3}$ and $v_{u_{2,3}}$, leading to analogous hierarchies for the
corresponding soft squared masses\footnote{In fact, the hierarchy is
  stronger in the $H_u$ sector since the ratio $v_{u_{2,3}}/v_{u_1}$
  must be smaller in order to explain the mass hierarchy between the
  top quark mass and the charm and up quark masses.}. Finally, as
discussed in section \ref{SBRPV}, the sneutrino VEVs are naturally
small since these \emph{``induced VEVs''} are proportional to the
$\epsilon$ parameter. This can be seen in the corresponding tadpole
equations,
\begin{eqnarray}
\frac{\partial V}{\partial v_{L_i}} = && \frac{1}{8} \Big( 8 \: \text{Re}(B_\epsilon) v_{u_i} + v_{L_i} (g_1^2+g_2^2)(|\vec v_d|^2 - |\vec v_u|^2 + |\vec v_L|^2) \nonumber \\
&& + 4 (m_{l_i l_i}^2 v_{L_i} + \sum_j m_{l_i l_j}^2 v_{L_j}) + 8 v_{L_i} |\epsilon|^2 - 4 \sqrt{2} v_{d_i} v_s \: \text{Re}(\epsilon \lambda^*) \Big)
\end{eqnarray}
which imply that all $v_{L_i}$ vanish in the $\epsilon = B_\epsilon = 0$ limit.

The requirement of very large soft squared masses has important
consequences on the mass spectrum. Dominated by the contributions from
the soft terms, the spectrum contains several degenerate
$\{H^0,A^0,H^\pm\}$ sets, some of them with masses in the $10-100$ TeV
range. This degeneracy is very strong in the case of $H^0$ and $A^0$
and only slightly broken for $H^\pm$ due to its mixing with the
charged sleptons.

Another interesting feature of the spectrum is the decrease of the
mass of the scalar singlet for increasing $\tan \beta$, where
\begin{equation}
\tan \beta = \frac{v_{u_1}}{v_{d_1}}
\end{equation}
In fact, for $\tan \beta \sim 10$ one can easily find points in
parameter space with a light scalar singlet, $S_1$, in the range of
$100$ MeV - $20$ GeV. This result can be easily understood in a simple
limit. Neglecting the mixing with the doublet states, the leading
contribution to the squared mass of the scalar singlet is
\begin{equation} \label{scalar-singlet}
m_{S_1}^2 = \frac{\tan \beta \: v^2 (2 \lambda m_S + T_\lambda)}{\sqrt{2}(1+\tan^2 \beta)v_s}
\end{equation}
where $v^2 = (246 \text{GeV})^2 = 4 m_W^2/g^2$ is the usual
electroweak VEV. Eq. \eqref{scalar-singlet} shows that $m_{S_1}^2$
scales as $1 / \tan \beta$ for sufficiently large $\tan \beta$. This
naturally leads to $m_{S_1}^2 \ll v^2$ for $\tan \beta \gtrsim 10$. In
such scenarios the Higgs decay channel $h \to S_1 S_1$ can dominate
the Higgs decay if the $h-S_1-S_1$ coupling is large. After the first
hints of the existence of the Higgs boson, later confirmed in
Refs.~\cite{CMSdiscovery,ATLASdiscovery}, it has been increasingly
clear that the properties of the discovered particle are very close to
those of a SM Higgs boson
\cite{Corbett:2012dm,Ellis:2012hz,Carmi:2012in,Plehn:2012iz}. Although
there is still room for new interactions \cite{Banerjee:2012xc}, these
are now constrained by the data. This imposes an important restriction
on the size of our $\lambda$ coupling, $\lambda \ll 1$. 


As a generic example to illustrate these properties, we provide the
following parameter set and results for a particular but generic point
in parameter space.

\begin{enumerate}
{\bf \item Parameters set}
\end{enumerate}
\begin{itemize}

\item Superpotential and $V_{\text{soft}}^{A_4}$ parameters: $\tan
  \beta = 30$, $\lambda = 0.01$, $m_S = 88$ TeV, $T_\lambda =
  -2.3$ TeV, $B_{m_S} = -0.79$ TeV$^2$, $B_\epsilon = -4.25$ TeV$^2$
  and $m_s^2 = -31000$ TeV$^2$.

\item $V_{\text{soft}}^{\slashed{A_4}}$ parameters:

\begin{eqnarray}
m_{H_{d_i} H_{d_j}}^2 &=& \left( \begin{array}{c c c}
4.4 & -9.9 & 2.2 \\
 & 500.0 & 5.1 \\
 & & 111.7
\end{array} \right) \: \text{TeV}^2 \\
m_{H_{u_i} H_{u_j}}^2 &=& \left( \begin{array}{c c c}
-0.003 & -7.6 & 6.2 \\
 & 37900.0 & -8.6 \\
 & & 31200.0
\end{array} \right) \: \text{TeV}^2 \\
m_{l_i l_j}^2 &=& \left( \begin{array}{c c c}
6800000.0 & 2.2 & -2.6 \\
 & 62200.0 & -9.6 \\
 & & 8300.0
\end{array} \right) \: \text{TeV}^2 
\end{eqnarray}

\item Neutrino physics can be accommodated with $\epsilon \sim
  10^{-4}$ GeV, which in turn results in $v_{L_i}$ of the same
  order. Similar soft terms are given in the charged scalar sector. In
  this parameter point one finds the VEV configuration: $v_{d_1}^2 +
  v_{u_1}^2 \simeq v^2$, $v_{d_2} = - v_{d_3} \sim 10^{-2} v_{d_1}$,
  $v_{u_2} = - v_{u_3} \sim 10^{-4} v_{u_1}$, and $v_s = -8$ TeV.

\end{itemize}

\begin{enumerate}
\setcounter{enumi}{1}
{\bf \item Results in the scalar sector}
\end{enumerate}
\begin{itemize}

\item A light scalar, $m_{SS} = 120$ MeV, of singlet nature with tiny
  ($\sim 10^{-8} \, \%$) doublet admixture.
\item A light scalar, $m_{h^0} = 90.4$ GeV, of doublet nature. This
  state can be identified with the Higgs boson~\footnote{For
    simplicity we give tree-level results. Large 1-loop corrections
    are of course expected and bring the mass of the light doublet
    state to experimentally acceptable levels.}.
\item Degeneracies between real and imaginary components of the
  sneutrinos.
\item Very heavy (and degenerate) states with masses $m_{H^0} \simeq
  m_{A^0} \simeq m_{H^\pm}$ in the multi-TeV range.

\end{itemize}
With this generic prediction for the spectrum in the extended scalar
sector the model turn outs to be safe from the constraints on the
oblique S, T, U parameters. Besides the heaviness of some states, the
degeneracies among their masses cancel their contributions to these
precision observables.
Note also that in our scenario the large supersymmetry breaking scale
is related to the flavor symmetry used to get the required vaccuum
alignment.

\black
\section{Discussion and conclusions}
\label{conclu}

We have extended the (next to) MSSM by implementing a discrete
non-Abelian flavor symmetry $A_4\times Z_2$. The most general
renormalizable allowed superpotential forbids all the trilinear RPV
terms (including those violating baryon number) and has a single
bilinear R-parity violating term.
Three copies of up and down Higgs doublets are required, in addition
to a \321times singlet present in $\hat{S}$, odd under $Z_2$.
When these develop VEVs both the electroweak and flavor symmetries are
broken and, in addition, sneutrinos acquire tiny VEVs. The Higgs
fields align so as to recover the correct charged fermion mass
hierarchies and the two successful predictions Eqs.~(\ref{eq:ql}) and
(\ref{Uf})~\cite{Morisi:2011pt}.

As in the usual flavor-less BRPV model, the 1-loop radiative
corrections is misaligned with the tree-level ones. These 1-loop
contributions provide the solar mass square splitting.
Due to the flavor symmetry and vacuum alignment, one neutrino is
nearly massles and there is a non-trivial restriction upon the
neutrino oscillation parameters, displayed in
Figs.~\ref{correlation1}-\ref{correlation3}, consistent however with
the most recent experimental data presented at the recent Neutrino
2012 conference and the corresponding neutrino oscillation global
fits~\cite{Tortola:2012te,Fogli:2012ua,Gonzalez}.
As far as collider physics is concerned, our model predicts that LSP
decays and neutrino mixing angles are tightly correlated, opening
encourageing expectations for searches at the
LHC~\cite{DeCampos:2010yu}.
Finally, even though the usual neutralino LSP is lost as dark matter
candidate, one can show that a relatively light gravitino provides a
very interesting alternative~\cite{Restrepo:2011rj}.

\acknowledgments

We thank Martin Hirsch for useful discussions.  Work supported by the
Spanish MEC under grants FPA2008-00319/FPA, FPA2011-22975 and
MULTIDARK CSD2009-00064 (Consolider-Ingenio 2010 Programme), by
Prometeo/2009/091 (Generalitat Valenciana), by the EU ITN UNILHC
PITN-GA-2009-237920. S. M. is supported by a Juan de la Cierva
contract, E. P.  by CONACyT (Mexico) and A.V. by the ANR project
CPV-LFV-LHC {NT09-508531}.

\end{document}